\documentclass[]{article}
\usepackage[T1]{fontenc}
\usepackage[utf8]{inputenc}
\usepackage[english]{babel}
\usepackage{amsmath}
\usepackage{amssymb}
\usepackage{graphicx}
\usepackage{multirow}
\usepackage{xspace}
\usepackage{fixltx2e}
\usepackage{paralist}
\usepackage{listings}
\usepackage{url}
\usepackage{booktabs}
\usepackage[pagebackref=false,breaklinks=true,letterpaper=true,colorlinks,bookmarks=false]{hyperref}
\usepackage[numbers]{natbib}
\usepackage[table]{xcolor}

\makeatletter

\pdfpageheight\paperheight
\pdfpagewidth\paperwidth

\graphicspath{{figures/}}

\newcommand{\utw}{\textsuperscript{1}}
\newcommand{\dcu}{\textsuperscript{2}}

\title{Adapting Binary Information Retrieval Evaluation Metrics for Segment-based Retrieval Tasks}

\author{
Robin Aly\utw \and
Maria Eskevich\dcu \and
Roeland Ordelman\utw \and
Gareth J.F. Jones\dcu \\
\utw University of Twente, The Netherlands\\
\dcu CNGL Centre for Global Intelligent Content, \\ 
School of Computing, Dublin City University, Ireland \\
}

\begin{document}

\date{10th December 2013}

\maketitle

\begin{abstract}
This report describes metrics for the evaluation of the effectiveness of segment-based retrieval based on existing binary information retrieval metrics. This metrics are described in the context of a task for the hyperlinking of video segments. This evaluation approach re-uses existing evaluation measures from the standard Cranfield evaluation paradigm. Our adaptation approach can in principle be used with any kind of effectiveness measure that uses binary relevance, and for other segment-baed retrieval tasks. In our video hyperlinking setting, we use precision at a cut-off rank $n$ and mean average precision.
\end{abstract}

\section{Introduction} 
\label{sec:introduction}

Video hyperlinking is an emerging research field~\cite{Eskevich2013a}. A popular and robust way to measure the quality of information retrieval systems and to compare them against each other is to use the Cranfield paradigm that requires four components: a collection of documents, a set of queries, a set of relevance judgments between documents and queries, and evaluation measures that should reflect the achieved effectiveness of results for users~\cite{Voorhees2009}. A difference between the video hyperlinking setting and traditional applications of the paradigm is that the documents in the collection are not predefined and retrieval systems can return segments of arbitrary start and length. Nevertheless, using established evaluation measures in video hyperlinking has the benefit of inheriting their established correlation with user effectiveness~\cite{Sanderson2010}. This paper describes a method of using established evaluation measures, such as precision at a certain rank or average precision, with adjustment to the varying video segmentation boundaries in the results.

\citet{Vries2004} were among the first to address the evaluation of retrieval systems that return arbitrary segments in XML retrieval and video search. Their key assumption is that users are likely to tolerate reviewing a certain amount of non-relevant content in a retrieved item before arriving at the beginning of the actual relevant content. They use this assumption to derive new evaluation measures. In this paper we make similar assumptions, but we aim to adapt existing evaluation measures because they are are widely understood and applied.

Note that an alternative to adapting the evaluation data to existing measures is to adapt the existing measures to the evaluation data, which we for example proposed in \cite{DBLP:conf/ecir/EskevichMJ12}.

The rest of this paper is structured as follows: Section~\ref{sec:usermodel} describes three alternative methods of incorporating users tolerance to non-relevant content into existing evaluation measures; Section~\ref{sec:implementation} gives details of our implementation; and Section~\ref{sec:conclusions} concludes the paper.

\section{User Models} 
\label{sec:usermodel}

\newcommand{\V}{\mathcal{V}}
\newcommand{\resv}{\vec{res}}
\newcommand{\res}{res}
\newcommand{\SIZE}{BS}

We begin our discussion by formalizing the existing Cranfield evaluation paradigm and the video hyperlinking evaluation scenario. Evaluation measures in the cranfield paradigm are functions of a string of relevance values from a domain $\V$ (we consider binary relevance where $\V=\{0,1\}$) of length $l$ to a value in the interval $[0:1]$: $m: \V_1\times ...\times \V_l \rightarrow [0:1]$, where the $i$th value corresponds to the relevance value of the result at rank $i$. The relevance values are obtained through relevance judgments $r(q,d)\rightarrow \V$ between a query $q$ and a document $d$, where $r(q,d)=1$ if the $d$ is relevant to $q$ and $0$ otherwise. For example, for a system result $\resv = d_1, ... d_n$, where $d_i$ is the $i$-th result, the evaluation measure precision at $n$, $m=P@n$ for a query $q$ can be defined as follows:
\[
P@n(\resv) = \frac{1}{n} \sum_{i=1}^n{r(q,d_i)}
\]
In video hyperlinking, relevance judgments and individual results are segments $\res=(d,s,e)$ where the document $d$ is a video, $s$ is the recommended start time and $e$ is the end time of the segment (with $s<e$). The relevance judgments can be seen as the definition of a function $r(q, \res)\rightarrow \{0,1\}$. We write a result list of size $n$ as $\resv = \res_1, ... , \res_n = (d_1, s_1, e_1),..., (d_n, s_n, e_n)$. The main difference between the Cranfield paradigm and video hyperlinking evaluation is that in the former case the relevance judgments $r$ are assumed to be easily defined for clearly defined document units, while variations in the start or end time of retrieved items in the latter case mean that the situation is much more complicated.

In order to use established evaluation measures, we propose three alternative adaptations to the relevance judgment function $r$.

\begin{figure}[h]
\includegraphics[width=1.0\textwidth]{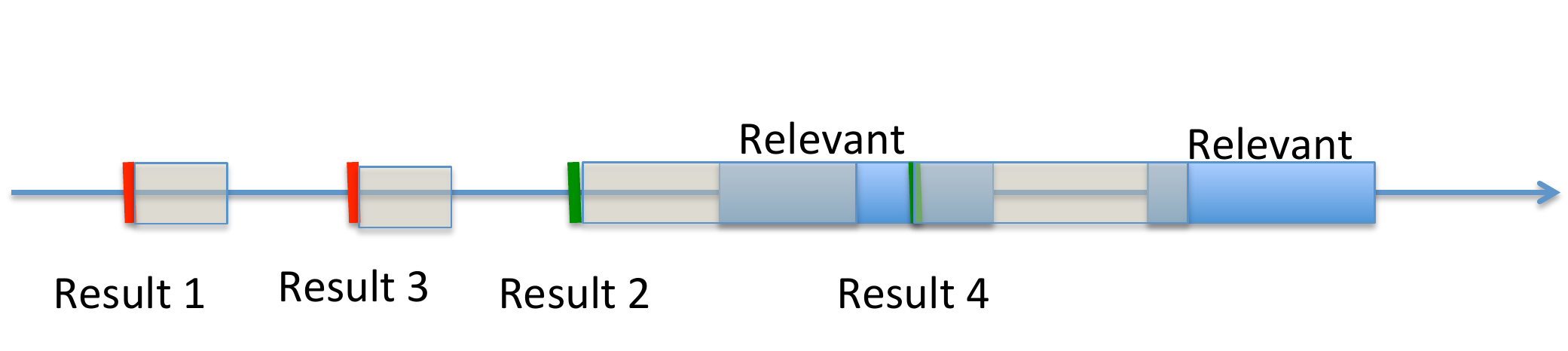}
\caption{Overlap Relevance: segments are relevant if they overlap with a relevant segment. \label{fig:overlap}}
\end{figure}

\subsection{Overlap Relevance}

Let $R_q = \{(d,s,e),... \}$ be the set of segments that were judged relevant for a query $q$, see Figure~\ref{fig:overlap} for an example. The overlap relevance alternative defines the relevance of a segment $\res$ by the fact of whether it overlaps with a relevant segment
\[
r(q,\res) := \res \in R_q
\]
where $\in$ is a binary operator for temporal overlap with one of the set members in $R_q$. An advantage of this alternative is that it can be implemented easily. However, it has the disadvantage that multiple result segments that overlap with a single relevant segment are counted multiple times, which is the case for result $2$ and result $4$ in Figure~\ref{fig:overlap}.

\begin{figure}[h]
\includegraphics[width=1.0\textwidth]{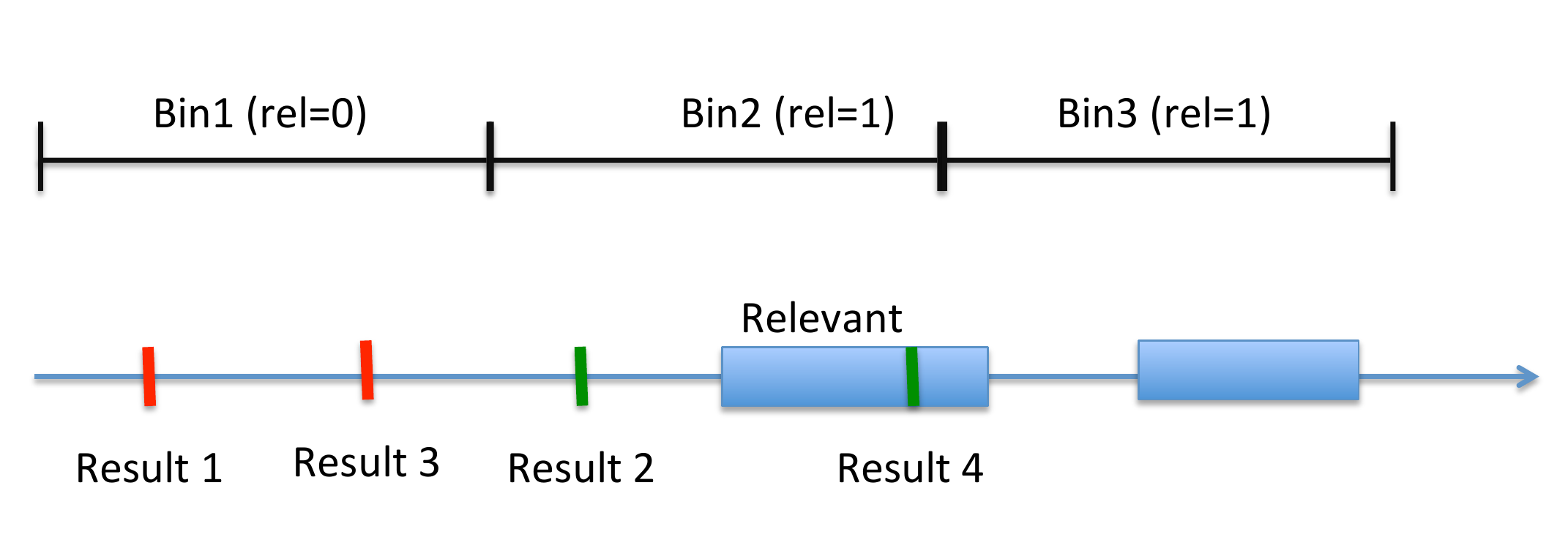}
\caption{Binned Relevance: relevant segments are put into bins, as well as result start times. A result segment is deemed relevant, if there is a relevance judgment in the bin the start time of the segment fits into. \label{fig:binned}}
\end{figure}

\subsection{Binned Relevance}

The binned relevance approach defines the result segments to units of a fixed size, which we refer to as bins, see Figure~\ref{fig:binned} for an example.

Let $R_q = \{(d,s,e),... \}$ be the set of segments that were judged relevant for a query $q$. Then we define an amended relevance set $R'_q=\{(d,s,e)| (d,s',e') \in R_q, s=lower(s', \SIZE), e=upper(s',\SIZE)\}$ where $lower$ returns the next smaller multiple of $\SIZE$ for start time $s'$ and $upper$ returns the next bigger multiple of $\SIZE$. For a given result list $\resv=(d_1, s_1, e_1),..., (d_n, s_n, e_n)$ the binned result list is: $\resv'=(d_1, lower(s_1, \SIZE), upper(s_1, \SIZE))$.

In the example in Figure~\ref{fig:binned}, bins $2$ and $3$ are considered as relevant because they contain at least one passage of relevant content, while bin $1$ does not contain any. Results $1$ and $3$ are merged into a single result in the ranked results list as they fit into the same bin $1$, thus the merged result is non relevant. The same procedure is carried out for the results $2$ and $4$ that are merged into bin $2$, which is judged as relevant.

\begin{figure}[h]
\includegraphics[width=1.0\textwidth]{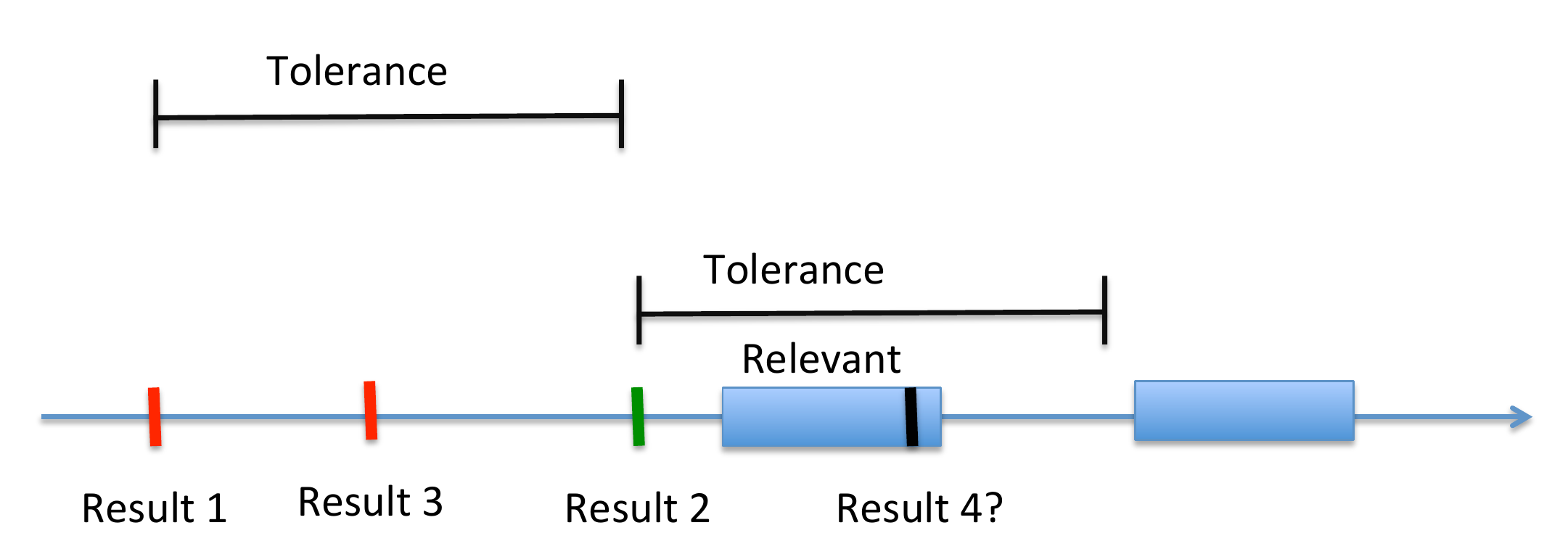}
\caption{Tolerance to Irrelevance: only the start times of segments are considered. \label{fig:tolerance}}
\end{figure}

\subsection{Tolerance to Irrelevance}

In the tolerance to irrelevance approach we modify the relevance judgment function $r$, which is shown in Figure~\ref{fig:tolerance}. The approach assumes that, given a result segment $(d,s,e)$ the user watches a fixed length of $l$ time units. If the start time of the relevant content is encountered by the user within the $l$ time units, the result is taken to be relevant. The redefinition of the relevance judgment function is: $r(q,(d,s,e)) = (d,s,s+l) \in R_q \cap (d,s,s+l) \not\in seen$ where $\in$ stands for temporal overlap, and $seen$ is the set of previously seen segments.

In the example in Figure~\ref{fig:tolerance}, result $1$ is non relevant because the user does not encounter relevant content within $l$ time units. Result $2$ is counted as relevant as it covers a relevant segment, whereas result $3$ is non relevant because the relevant content has been already seen by the user, when the examined result $2$. Similarly, result $4$ is counted as non relevant because it has already been seen through result $2$.

\section{Implementation} 
\label{sec:implementation}

We provide a script that calculates the above measures. The script is called as follows:
\begin{verbatim}
  me13sh_eval.py qrel ranking
\end{verbatim}
where qrel is a file containing relevance judgment and ranking is a file containing the results of a system.

The output of the script is similar to the treceval script~\cite{treceval} and consists of three tab separated  columns where the first indicates the type of measure, the second indicates whether the measurement was performed for a particular query or link, and the third column contains the actual measurement value. The following is a partial example of the output. 

\begin{verbatim}
num_q            	all      	30    
videos_ret       	all      	19    
videos_rel       	all      	27    
avglength_ret    	all      	119   
avglength_rel    	all      	104   
num_rel          	all      	1673  
num_ret          	all      	2984  
num_rel_ret      	all      	789   
map              	all      	0.3000
P_5              	all      	0.7000
P_10             	all      	0.6567
P_20             	all      	0.5450
Judged_10        	all      	1.0000
Judged_20        	all      	0.7933
Judged_30        	all      	0.6789
num_rel_bin      	all      	1514  
num_ret_bin      	all      	1896  
num_rel_ret_bin  	all      	421   
map_bin          	all      	0.1594
P_5_bin          	all      	0.6000
P_10_bin         	all      	0.5600
P_20_bin         	all      	0.4033
Judged_10_bin    	all      	0.8967
Judged_20_bin    	all      	0.6333
Judged_30_bin    	all      	0.5222
num_rel_tol      	all      	1673  
num_ret_tol      	all      	2984  
num_rel_ret_tol  	all      	348   
map_tol          	all      	0.0997
P_5_tol          	all      	0.5333
P_10_tol         	all      	0.4533
P_20_tol         	all      	0.3217
Judged_10_tol    	all      	1.0000
Judged_20_tol    	all      	0.7650
Judged_30_tol    	all      	0.6422
\end{verbatim}

Here, the measures ending with \_bin and \_tol refer to the binned relevance and the tolerance to irrelevance alternative respectively. The keyword all in the second column indicates that the value in the third column is an average over all items listed above. P\_$n$ is the precision at $n$ measure and Judged\_$n$ is a measure of how many segments have been judged within the top-$n$. 

\section{Concluding Remarks} 
\label{sec:conclusions}

We have presented three alternative adaptations of existing IR evaluation metrics from the document-based Cranfield paradigm for the segment-based video-hyperlinking task. The three alternatives are: relevance overlap, binned relevance, and tolerance to irrelevance. Simple overlap count a document as relevant if it overlap with a relevant segment. The binned relevance segment cluster temporally close segments together, and therefore focus on measuring the general ability of an algorithm to recommend segments in relevant regions. The tolerance to irrelevance approach assumes that a user watches a fixed amount of time from a given start point. If the user encounters relevant content from the start point of a returned segment, this segment is counted as relevant.

\bibliographystyle{abbrvnat}
\bibliography{robib}
\end{document}